\documentclass[twocolumn,twoside,slac_two]{revtex4}
\usepackage{graphicx}
\usepackage{fancyhdr}
\usepackage{xspace}
\usepackage{relsize}
\def\babar{\mbox{\slshape B\kern-0.1em{\smaller A}\kern-0.1em
    B\kern-0.1em{\smaller A\kern-0.2em R}}}

\def\epem   {$e^+ e^-$}
\def\FourS {$\Upsilon(4S)$}
\def\invfb   {${\rm fb}^{-1}$}
\def\invpb   {${\rm pb}^{-1}$}
\def\invab   {${\rm ab}^{-1}$}

\def\piz   {$\pi^0$}

\def\pip   {$\pi^+$}
\def\pim   {$\pi^-$}

\def\pipm  {$\pi^\pm$}
\def\pimp  {$\pi^\mp$}

\def\Dbar    {$\kern 0.2em\overline{\kern -0.2em D}$}
\def\Db      {$\kern 0.2em\overline{\kern -0.2em D}$}
\def\Dz      {$D^0$}
\def\Dzb     {$\kern 0.2em\overline{\kern -0.2em D}^0$}
\def\DzDzb   {$D^0 {\kern -0.16em \kern 0.2em\overline{\kern -0.2em D}}$}

\def\Dm      {$D^-$}

\def\DpDm    {$D^+ {\kern -0.16em \Dm}$}

\def\Dstarb  {$\kern 0.2em\overline{\kern -0.2em D}^*$}

\def\Dstarzb {$\kern 0.2em\overline{\kern -0.2em D}^{*0}$}

\def\Dsb     {$\kern 0.2em\overline{\kern -0.2em D}^+_s$}

\def\dodstar {$D^{(*)0}$}
\def\dodstarb {$\kern 0.2em\overline{\kern -0.2em D}^{(*)0}$}
\def\Dztilde   {$\tilde{D}^0$}
\def\Dstarztilde   {$\tilde{D}^{\ast 0}$}

\def\Kbar  {$\kern 0.2em\overline{\kern -0.2em K}$}

\def\Kzb   {$\Kbar^0$}
\def\KzKzb {$K^0 \kern -0.16em \Kzb$}
\def\Kp    {$K^+$}
\def\Km    {$K^-$}

\def\KpKm  {$K^+ \kern -0.16em K^-$}
\def\KS    {$K^0_{\scriptscriptstyle S}$}

\def\Kstarzb {$\kern 0.2em\overline{\kern -0.2em K}^{*0}$}

\def\Kstarb  {$\kern 0.2em\overline{\kern -0.2em K}^*$}

\def\Kstarm  {$K^{*-}$}

\def\btodtildek   {$\tilde{D}^0 K^-$}
\def\btodsttildek   {$\tilde{D}^{*0} K^-$}
\def\btodtildekst   {$\tilde{D}^0 K^{*-}$}
\def\btodtildekstzero   {$\tilde{D}^{(\ast) 0} \bar{K}^{(\ast) 0}$}

\def\btodtildekpm   {$\tilde{D}^0 K^\pm$}
\def\btodsttildekpm   {$\tilde{D}^{*0} K^\pm$}
\def\btodtildekstpm   {$\tilde{D}^0 K^{* \pm}$}

\def\KpKm  {$\Kp \kern -0.16em \Km$}

\def\Bp      {$B^+$}
\def\Bm      {$B^-$}

\def\BpBm    {$B^+ {\kern -0.16em B^-}$}
\def\Bbar    {$\kern 0.18em\overline{\kern -0.18em B}$}
\def\BB      {$B \kern 0.18em\overline{\kern -0.18em B}$}
\def\Bz      {$B^0$}
\def\Bzb     {$\kern 0.18em\overline{\kern -0.18em B}^0$}

\def\cpp    {$CP_{+}$}
\def\cpm    {$CP_{-}$}

\def\Acp    {${\mathbf A}_{CP }$}
\def\Acpp   {${\mathbf A}_{CP_{+}}$}
\def\Acpm   {${\mathbf A}_{CP_{-} }$}
\def\Acppm  {${\mathbf A}_{CP \pm }$}
\def\Rcp    {${\mathbf R}_{CP }$}
\def\Rcpp   {${\mathbf R}_{CP_{+} }$}
\def\Rcpm   {${\mathbf R}_{CP_{-} }$}
\def\Rcppm  {${\mathbf R}_{CP \pm }$}

\def\AADS    {${\mathbf A}_{ADS }$}
\def\RADS    {${\mathbf R}_{ADS }$}

\newcommand{\nimBaseA}       {Nucl.\ Instr.\ Methods Phys.\ Res., Sect.\ A\xspace}
\newcommand{\nima}      [1]  {\nimBaseA~{\bf #1}}
\newcommand{\jplBase}        {Phys.\ Lett.\xspace}
\newcommand{\plb}       [1]  {\jplBase\ B~{\bf #1}}
\newcommand{\jprBase}        {Phys.\ Rev.\xspace}
\newcommand{\jprd}      [1]  {\jprBase\ D~{\bf #1}}
\newcommand{\jprlBase}       {Phys.\ Rev.\ Lett.\xspace}
\newcommand{\jprl}      [1]  {\jprlBase\ {\bf #1}}

\begin{document}

\title{Measurements of the CKM angle $\phi_{3}/\gamma$}

%

\author{V. Tisserand, for the \babar\ and Belle collaborations. }
\affiliation{Laboratoire de Physique des Particules (LAPP),
IN2P3/CNRS et Universit\'e de Savoie,\\  F-74941 Annecy-Le-Vieux,
France}

\begin{abstract}
We present a review on the measurements of the CKM angle $\gamma$
($\phi_3$)\footnote{The \babar\ and Belle collaborations use
different conventions to label the three CKM angles, where
$\beta\equiv \phi_1$, $\alpha \equiv \phi_2$, and $\gamma \equiv
\phi_3$. In this report we use the \babar\ convention: $\alpha$,
$\beta$, and $\gamma$.} as performed by the \babar\ and Belle
experiments at the asymmetric-energy \epem\ $B$ factories
colliders PEP-II and KEK$B$. These measurements are using either
charged or neutral $B$ decays. For charged $B$ decays the modes
{\btodtildek}, \btodsttildek, and \btodtildekst\ are employed,
where \Dztilde\ indicates either a \Dz\ or a \Dzb\ meson. Direct
{\it CP} violation is exploited. It is caused by interferences
between $V_{ub}$ and $V_{cb}$ accessible transitions that generate
asymmetries in the final states. For these decays various methods
exist to enhance the sensitivity to the $V_{ub}$ transition,
carrying the weak phase $\gamma$. For neutral $B$ decays, the
modes $D^{(*)\pm}\pi^{\mp}$ and $D^{\pm}\rho^{\mp}$ are used. In
addition to the $V_{ub}$ and $V_{cb}$ interferences, these modes
are sensitive to the \Bz-\Bzb\ mixing, so that time dependent
analyses are performed to extract $\sin(2\beta+\gamma)$. An
alternative method would use the lower branching ratios decay
modes \btodtildekstzero\ where much larger asymmetries are
expected.

The various available methods are mostly ``theoretically clean"
and always free of penguins diagrams. In some cases a high
sensitivity to $\gamma$ is expected and large asymmetries may be
seen. But these measurements are always experimentally difficult
as one has to face with either low branching ratios, or small
asymmetries, or additional technical/theoretical difficulties due
to Dalitz/SU(3) and re-scattering models needed to treat/estimate
nuisance parameters such as unknown strong phases and the relative
magnitude of the amplitude of the interfering ``$V_{ub}$"
transitions. Thus at the present time only a relatively limited
precision on $\gamma$ can be extracted from these measurements.
The current world average is
$\gamma=(78^{+19}_{-26})^{\circ}$~\cite{CKMFitter}. For other
methods and long term perspectives, as discussed in details, the
reader is invited to consult the proceedings of the recent CKM
workshop that was held in Nagoya (Japan) in December
2006~\cite{CKM2006}.
\end{abstract}

\maketitle

\thispagestyle{fancy}

\section{Introduction}
We present the results of the measurements performed by the
\babar\ and Belle collaborations, to determine the value of the
Cabibbo-Kobayashi-Maskawa (CKM) {\it CP} violating phase $\gamma$
($\equiv\arg{\left[-V_{ud}V_{ub}^{*}/V_{cd}V_{cb}^{*}\,\right]}$).

 These measurements are based on the studies done with the
charged $B$ decays \btodtildek, \btodsttildek, and \btodtildekst,
where \Dztilde\ indicates either a \Dz\ or a \Dzb\ meson. They are
based on the interference of the amplitudes proportional to the
$V_{cb}$ and $V_{ub}$ CKM-matrix elements that generate
asymmetries through direct {\it CP} violation effects associated
to the electroweak ({\it EW}) phase $\gamma$ carried by the
$V_{ub}$ amplitude. These measurements are at the present time the
most constraining ones, but neutral $B$ decays such as
$D^{(*)\pm}\pi^{\mp}$ and $D^{\pm}\rho^{\mp}$ are also used to
extract constraints on $\sin(2\beta+\gamma)$, as \Bz-\Bzb\ mixing
is present in addition to the above described direct {\it CP}
violation phenomenon. The sensitivity to the $V_{ub}$ amplitude is
relatively small for these decays, so it has been proposed to use
the rare decays \btodtildekstzero\ where larger {\it CP}
asymmetries are expected.

At the time of this conference the two  asymmetric-energy \epem\
colliders PEP-II at SLAC and KEK$B$ at KEK have produced a huge
quantity of data at the \FourS\ resonance. \babar~\cite{babar} and
Belle~\cite{belle} detectors have integrated over $1100$~\invfb\
of data (respectively about $420$~\invfb\ and $710$~\invfb). This
corresponds to a sample of more than one billion \BB\ pairs
collected. It can be noticed that the measurements presented here
are all statistically limited and most of them use only a fraction
of the currently available dataset. It is therefore crucial that
they will be updated soon.

\section{Measuring $\gamma$ with charged $B$ decays}

Methods are exploited~\cite{GLW,ADS,GGSZ}, where the \Dztilde\
decays either to a {\it CP} eigenstate ({\it GLW} method), or to a
Doubly Cabibbo-Suppressed  flavor Decay (DCSD, ``wrong sign"
decay, {\it ADS} method), or to the \KS \pim \pip\  final state,
for which a Dalitz analysis has to be performed ({\it GGSZ}
method). To extract $\gamma$, those three methods are all based on
the fact that a \Bm\ can decay into a color-allowed \Dz \Km\
(color-suppressed \Dzb \Km) final state via $b \to c \bar{u} s$
($b \to u \bar{c} s$) transitions. The amplitude ${\cal
A}(``V_{cb}")$ of the $b \to c \bar{u} s$ transition is
proportional to $\lambda^3$ and the amplitude ${\cal
A}(``V_{ub}")$ of the $b \to u \bar{c} s$ transition to $\lambda^3
\sqrt{\bar{\eta}^2+\bar{\rho}^2} e^{i( \delta_B - \gamma)}$ (where
$\lambda$ is the related to the Cabibbo angle). The second
amplitude therefore carries both the {\it EW} $\gamma$ {\it CP}
phase and the relative strong phase of those two transitions. As
the total amplitude for the \btodtildek\ decay is the sum of the
two amplitudes ${\cal A}(`` V_{cb}")$ and ${\cal A}(``V_{ub}")$,
the two amplitudes interfere when the \Dz\ and \Dzb\ decay into
the same final state. This interference leads to different \Bp\
and \Bm\ decay rates (direct {\it CP} violation). These methods
apply to the two other charged $B$ decays: \btodsttildek and
\btodtildekst, as well.

The various  methods are ``theoretically clean" because the main
contributions to the amplitudes come from tree-level transitions,
in an excellent approximation. In addition to the parameters and
to the strong phase, ${\cal A}(``V_{ub}")$ is significantly
reduced with respect to ${\cal A}(``V_{cb}")$ by the
color-suppression phenomenon. One usually defines the parameter
${\rm r}_B \equiv \vert {\cal A}(``V_{ub}")/{\cal A}(``V_{cb}")
\vert$ that determines the size of the direct {\it CP} asymmetry.
It is a critical parameter for these analyses. Its value is
predicted~\cite{gronau} to lie in the range $0.05-0.3$ as it has
been argued that color-suppression may not be as important as
expected from naive factorization in charmed B
decays~\cite{BaBBeColSup}. The smaller ${\rm r}_B$ is, the smaller
is the experimental sensitivity to $\gamma$.

As the three decays modes \btodtildek, \btodsttildek, and
\btodtildekst\ are used, one should remark that 7 parameters have
to be extracted from the three used methods: the common {\it EW}
phase $\gamma$ and the nuisance parameters for the measurements,
i.e., a respective strong phase ``$\delta_B$" and an ``${\rm
r}_B$" parameter for each of the three modes: $\delta_B$, ${\rm
r}_B$, $\delta^{*}_B$, ${\rm r}^{*}_B$, $\delta_{sB}$, and ${\rm
r}_{sB}$.

The CKM-angle $\gamma$, and the parameters ``${\rm r}_B$", and
``$\delta_B$" can be measured experimentally by two quantities
(Asymmetry and Ratio of branching ratios):

\begin{equation}
 {\mathbf{A}} \equiv \frac{ \Gamma{  (B^- \to { \tilde{
D}^{(*)0}} K^{(*)-})} {  {{ -}}} \Gamma{  (B^+ \to { \tilde{
D}^{(*)0}} K^{(*)+})} } {\Gamma{  (B^- \to { \tilde{ D}^{(*)0}}
K^{(*)-})} { +} \Gamma{(B^+ \to { \tilde{D}^{(*)0}} K^{(*)+})} }
\end{equation}

\begin{equation}
{\mathbf{R}} \equiv \frac{ \Gamma{  (B^- \to { \tilde{ D}^{(*)0}}
K^{(*)-})} { { { +}}} \Gamma{  (B^+ \to { \tilde{ D}^{(*)0}}
K^{(*)+})} } { \Gamma{ (B^- \to { D}^{(*)0} K^{(*)-})} + \Gamma{
(B^+ \to \bar{ D}^{(*)0} K^{(*)+})}}
\end{equation}

For ${\mathbf{R}}$, the denominator is  built with branching
ratios of specific flavor \Dz\ decays.

\subsection{The {\it GLW} method: \Dztilde\ CP eigenstates}
\label{section:GLW}

For the {\it GLW} method~\cite{GLW} the \Dztilde\  is
reconstructed in various {\it CP} eigenstate decay
channels~\cite{BelleGLW,BaBarGLW}: $K^+K^-$,  $\pi^+ \pi^-$ (\cpp\
eigenstates); and \KS\piz, \KS$\phi$, \KS$\omega$ (\cpm\
eigenstates). The \Dstarztilde\ is reconstructed in the decay mode
\Dztilde$\pi^0$ only and the $K^*$(892)$^-$ mesons into the decay
\Kstarm$\to$\KS\pim. The total branching ratio of each considered
decay mode, including secondary decays is relatively small
($10^{-6}$ or less). As many modes are taken into account to
reconstruct the \Dztilde\ it is anyway possible to have enough
signal events to study asymmetries, but this technique is
obviously strongly statistically limited. So the small {\it CP}
asymmetry (${\rm r}_B \simeq 0.05-0.3$) and the rareness of these
\Dz\ {\it CP} eigenstate modes, make this method difficult with
the present $B$ factories dataset.

In the {\it GLW} method, there exist 4 observable quantities, for
three unknowns ($\gamma$, ${\rm r}_B$, and $\delta_B$): \Rcppm $=
{1 + {\rm r}_{B}^{2} \pm 2 {\rm r}_{B} \cos \delta \cos \gamma }$
and \Acppm $= \pm \ 2 {\rm r}_{B} \sin \delta \sin \gamma
/$\Rcppm. Only three are independent, as: \Rcpm\Acpm=-\Rcpp\Acpp.
The observable \Rcp\ is normalized to the branching ratios as
obtained from three flavor state decays: \Dz $\to$  \Km\pip, \Km
\pip \piz, and \Km\pip\pip\pim. In principle with infinite
statistics this method is very clean to determine $\gamma$, but up
to an 8 fold-ambiguity, as the above equations show. It can be
noticed from the definition of \Rcp\ that the sensitivity to the
nuisance parameter ${\rm r}_{B}$ is quite weak, as it is expected
to be much smaller than unity.

\begin{figure}[h]
\centering
\includegraphics[width=80mm]{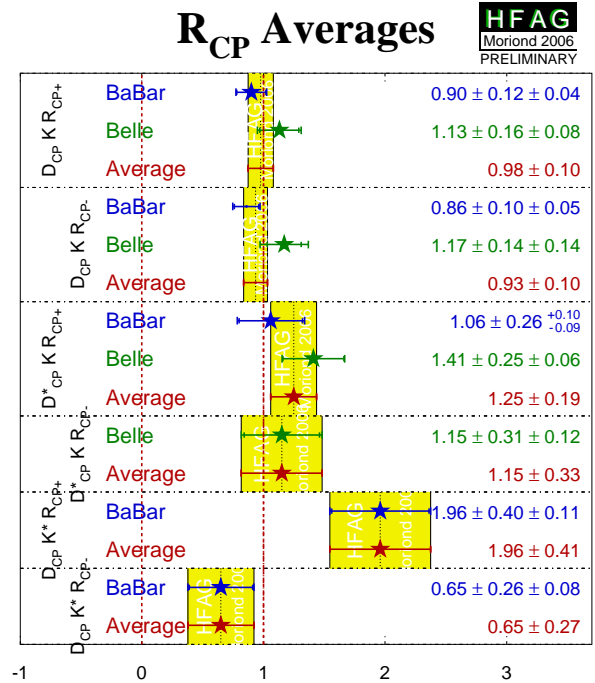}
\caption{World HFAG~\cite{HFAG} compilation on the \Rcp\
observable.} \label{fig:GLW_RCP}
\end{figure}

\begin{figure}[h]
\centering
\includegraphics[width=80mm]{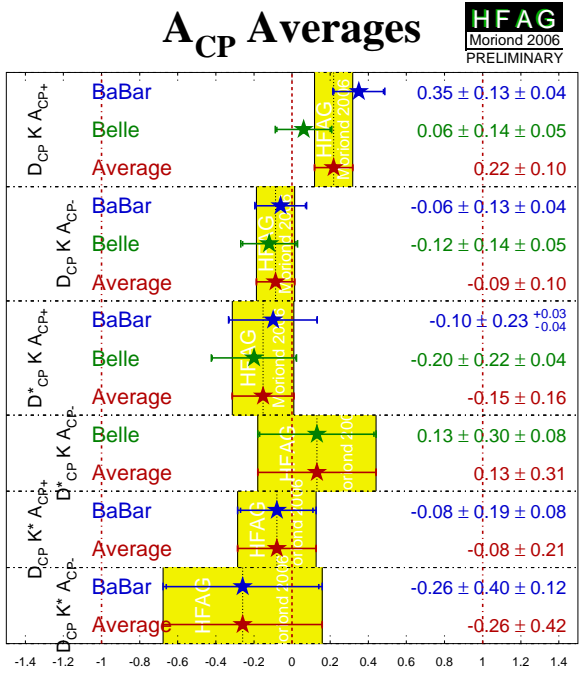}
\caption{World HFAG~\cite{HFAG} compilation on the \Acp\
observable.} \label{fig:GLW_ACP}
\end{figure}

The \babar~\cite{BaBarGLW} and Belle~\cite{BelleGLW}
collaborations have published results based on a limited fraction
only of their full available dataset. They respectively use $232
\times 10^6$ and $275 \times 10^6$ \BB\ pairs to measure the
observables \Rcp\ and \Acp. Significant signals allow to compute
\Rcp\ and \Acp\ observables with however limited precision. The
(peaking)-background is estimated from the $m_{ES}$ ($m_{bc}$) and
\Dz\ mass sidebands. The \cpp\ pollution for \cpm\ eigenstate from
decays $K^0_S[K^+ K^-]_{{\rm non} \ \phi}$ and
$K^0_S[\pi^+\pi^-\pi^0]_{{\rm non} \ \omega}$ is estimated using
data. Finally, in the systematic uncertainty accounting,  the
possible strong phases as generated by probable $K \pi$ $S$ waves
in the \Kstarm $\to$ \KS \pim\ decays are taken into account.

Figures~\ref{fig:GLW_RCP} and~\ref{fig:GLW_ACP} summarize the
averages computed by the HFAG collaboration~\cite{HFAG} for the
measurements of the two $B$ factories. At the present time, the
measured values of \Acp\ (\Rcp) are not precise enough to differ
significantly from zero (unity) so that a precise constraint on
$\gamma$ can not be obtained from the {\it GLW} method alone.

From \Rcppm\ and  \Acppm\ one can derive the so called {\it
Cartesian coordinates}: ${\rm x}_\pm \equiv {\rm
r}_{B}\,\cos(\delta_{B}\pm\gamma)$, as ${\rm x}_\pm = \left( {
{\mathbf R}_{CP+ } (1 \mp {\mathbf A}_{CP+ })- {\mathbf R}_{CP-} (
1 \mp {\mathbf R}_{CP+ })} \right) /{4}$ and ${\rm r}^2_B=\left(
{{\mathbf R}_{CP+ }+{\mathbf R}_{CP-}-2} \right) /{2}$.  For the
\btodtildek\ \babar~\cite{BaBarGLW} extracts ${\rm r}_{B}^2=-0.12
\pm 0.08 { (\rm stat)} \pm 0.03 {(\rm syst)}$, and ${\rm x}_+ =
-0.082 \pm 0.052 { (\rm stat)} \pm 0.018 { (\rm syst)}$, ${\rm
x}_- = 0.102 \pm 0.062 { (\rm stat)} \pm 0.022 {(\rm syst)}$ and
for \btodtildekst \babar\ obtains ${\rm r}_{sB}^2=0.30 \pm 0.25$,
${\rm x}_{s+} = 0.32 \pm 0.18 { (\rm stat)} \pm 0.07 {(\rm
syst)}$, and ${\rm x}_{s-} = 0.33 \pm 0.16 { (\rm stat)} \pm 0.06
{ (\rm syst)}$. These measurements have a precision already
competitive with those of the {\it GGSZ} method (see
Sect.~\ref{section:GGSZ}), that is the reason why the {\it GLW}
method is useful into the global fit to $\gamma$ in a combined
statistical treatment of the various methods.

\subsection{The {\it ADS} method: \Dztilde\ Doubly Cabibbo Suppressed Decays (DCSD)}
\label{section:ADS}

For the {\it ADS} method~\cite{ADS}, the \Dz\ meson as generated
from the $b \to c \bar{u} s$ transition is required to decay to
the Doubly Cabibbo-Suppressed $K^+\pi^-$ mode (DCSD or "wrong
sign"), while the \Dzb meson, from the $b \to u \bar{c} s$
transition, decays to Cabibbo-favored final state $K^+\pi^-$. The
overall branching ratio for a final state $B^- \rightarrow
[K^+\pi^-]_{\tilde{D}^0} K^{(*)-}$ is expected to be very small
($\sim 10^{-6}$), but the two interfering diagrams are of the same
order of magnitude. The challenge in this method is therefore to
detect $B$ candidate in this final state with two opposite charge
kaons. The total amplitude is complicated by an additional unknown
relative strong phase $\delta_D$ in the \Dz-\Dzb $\rightarrow
[K^+\pi^-]$ system, while the ratio of their respective amplitude
${\rm r}^2_D$ is now very precisely measured and is equal to
$(0.376\pm0.009)\%$~\cite{pdg2006}. It can be written as
$A([K^+\pi^-]_{\tilde{D}^0} K^{(*)-})\propto
 {\rm r}_Be^{i(\delta_B-\gamma)}+{\rm r}_D e^{-i\delta_D}$.
Using the $B^- \rightarrow [K^-\pi^+] K^{(*)-}$ modes as
normalisation for \RADS, one can write the equations for the two
experimental observable quantities: \RADS$={ {\rm r}^2_B} + {{
{\rm r}^2_D}}+ 2{{{ {\rm r}_B}} {{ {\rm r}_D}}\cos({{ \delta_B}+{
\delta_D}}) \cos ({ { \gamma}})} $ and \AADS$=  {2 \ {{ {\rm
r}_B}} {{ {\rm r}_D}} \sin({{ \delta_B}+{ \delta_D}}) \sin ({ {
\gamma}})} / $\RADS, where \RADS\ is clearly highly sensitive to
${\rm r}^2_B$. The observable \AADS\ can obviously only be
measured if a significant number of DCSD candidates is seen.

For the  \btodtildek\ and \btodsttildek\ channels~\cite{BaBarADS}
and ~\cite{BelleADS}, no significant signal has been measured yet.
For the \btodsttildek\ modes \babar\ uses both the $D^{*0} \to
D^0\pi^0$ and $D^0 \gamma$ modes. It has been
demonstrated~\cite{BomdarGershon} that the strong phase
$\delta^*_B$ differs by an effective phase $\pi$, so that they can
be combined to extract $\delta^*_B$ and to set a more constraining
limit on ${\rm r}^*_B$. These measurements have been obtained with
dataset corresponding respectively to $232 \times 10^6$ and $386
\times 10^6$ \BB\ pairs. At 90 \% of confidence level (C.L.),
\babar\ sets the upper limits ${\rm r}_B<0.23$ and ${\rm
r^*}^2_B<(0.16)^2$. With about 1.7 times the \babar\ statistics
Belle obtains a slightly better limit: ${\rm r}_B<0.18$ at 90 \%
of C.L.. For the \btodtildekst\ decay~\cite{BaBarADS}, no
significant signal is seen yet, \babar\ measures \RADS$= 0.046 \pm
0.031 ({\rm stat})  \pm 0.008 ({\rm syst}) $, \AADS$= -0.22\pm
0.61 ({\rm stat})  \pm 0.17 ({\rm syst}) $. As part of the
systematic uncertainty accounting, \babar\ considers effect of the
possible strong phases as generated by probable $K\pi$ $S$ waves
in the \Kstarm$\to$\KS\pim\ decays. It is the dominant
contribution.

Using a  frequentist approach~\cite{CKMFitter}, and combining both
the {\it GLW} and {\it ADS} methods for the \btodtildekst\
channel~\cite{BaBarADS}, \babar\ extracts ${\rm
r}_{sB}=0.28^{+0.06}_{-0.10}$, and excludes at the two-standard
deviation level the interval $75^\circ < \gamma < 105^\circ$.

More recently, \babar~\cite{BaBarADSNew} has performed a
measurement on another DCSD using $226 \times 10^6$ \BB\ pairs:
the \Dz\ ``wrong sign" decays to $K^+$\pim\piz. In this case the
extraction of $\gamma$ is complicated by the variation of the \Dz\
decay amplitude and of the strong phase $\delta_D$ over the Dalitz
decay plane $K^+$\pim\piz. The relative value of ${\rm r}^2_D$ is
smaller than for the $K^+$\pim\ decay and equals to $(0.241\pm
0.011)\%$~\cite{BaBarDCSDKpipi0}, a larger sensitivity on \RADS\
and ${\rm r}_B$ is therefore expected. The price to pay is however
a larger background level. From a fit to $\Delta E$, $m_{ES}$, and
a neural-network multi-variable  discriminant to suppress the
light $q\bar{q}$ pairs background, \babar\ extracts
$18^{+18}_{-15}$ candidates, compatible with no DCSD signal, and
sets the Bayesian limits: \RADS$<0.039$ and ${\rm r}_B<0.185$ at
95 \% of credibility interval. One can remark that this constraint
on ${\rm r}_B$ is at the same level of sensitivity as the one
obtained from the $K^+$\pim\ DCSD decay.

\begin{figure}[h]
\centering
\includegraphics[width=80mm]{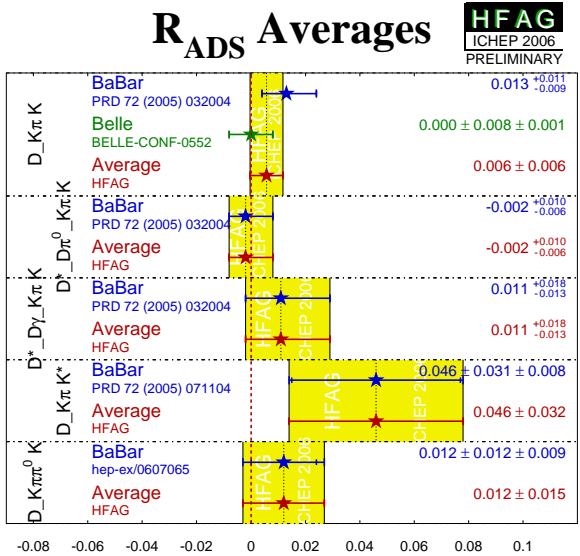}
\caption{World HFAG~\cite{HFAG} compilation on the \RADS\
observable.} \label{fig:RADS}
\end{figure}

Figure~\ref{fig:RADS} summarizes the averages computed by the HFAG
collaboration~\cite{HFAG} for the measurements of the two $B$
factories. At the present time, the measured values of \RADS\ are
not precise enough. So only limits on ``${\rm r}_B$" parameters
are set.  By extension the  {\it ADS} method can not provide us
yet with any strong constraint on $\gamma$ alone.

\subsection{The {\it GGSZ} method: \Dztilde\ to \KS \pim \pip\ Dalitz decay analysis}
\label{section:GGSZ}

Among the \Dztilde\ decay modes studied so far the \KS \pim \pip\
channel is the one with the highest sensitivity to
$\gamma$~\cite{GGSZ} because of the best overall combination of
branching ratio magnitude, \Dz-\Dzb\ interference and background
level. This mode offers a reasonably high branching ratio ($\sim
10^{-5}$, including secondary decays) and a clean experimental
signature (only charged tracks in the final state). The decay mode
\KS\pim\pip\ can be accessed through many intermediate states:
``wrong sign" or ``right" $K^*$ resonances (such as $K^*(892)$,
$K^*(1410)$, $K_0^*(1430)$, $K_2^*(1430)$, or $K^*(1680)$), {\it
CP} eigenstates \KS$\rho^0$, \KS$\omega$, or \KS $f^0$, ...
Therefore, an analysis of the amplitude of the \Dztilde decay over
the $m^2(K^0_S\pi^-)$ vs $m^2(K^0_S\pi^+)$ (i.e.: $m^2_-$ vs
$m^2_+$) Dalitz plane structure is sensitive to the same kind of
observable as for both the {\it GLW} and {\it ADS} methods, and
obviously carries more information. The sensitivity to $\gamma$
varies strongly over the Dalitz plane. The contribution from the
$b \to u \bar{c} s$ transition in the $B^- \to$ \dodstar
$K^{(*)-}$ ($B^+ \to$ \dodstarb $K^{(*)+}$) decay can
significantly be amplified by the amplitude ${\cal A}_{D+}$
(${\cal A}_{D-}$) of the \Dzb $\to$ \KS\pim\pip\ (\Dz $\to$
\KS\pip\pim) decay (${\cal A}_{D\mp} \equiv {\cal
A}_{D}(m^2_\mp,m^2_\pm)$). Assuming no {\it CP} asymmetry in $D$
decays, and neglecting  \Dz-\Dzb\ mixing, the decay  rate of the
chain $B^- \to$ \dodstar $K^{(*)-}$ ($B^+ \to$ \dodstarb
$K^{(*)+}$), and \Dztilde~$\to$ \KS\pim\pip, can be written as:

\begin{equation}
\begin{array}{rcl}
\Gamma_\mp(m^2_-,m^2_+)  & \propto & |{\cal A}_{D\mp}|^2
    + {\rm r}_B^2 |{\cal A}_{D\pm}|^2   \nonumber \\
   & + & 2 \{ {\rm x}_\mp \mathop{\rm Re}[{\cal A}_{D\mp} {\cal A}_{D \pm}^*] \\
   & + &   {\rm y}_\mp \mathop{\rm Im}[ {\cal A}_{D\mp} {\cal A}^*_{D\pm}] \}.
   \nonumber
   \end{array}
\end{equation}

In the above equation, the {\it Cartesian coordinates} have been
introduced: $ \left\{ {\rm x}_\pm,   {\rm y}_\pm  \right\} =  \{
\mathop{\rm Re} , \mathop{\rm Im}  \}[ {\rm r}_B e^{i(\delta_B \pm
\gamma)}]$, for which the constraint ${\rm r}_B^2={\rm x}_\pm^2
+{\rm y}_\pm^2$ holds. They replace the physical constants:
``$\delta_B$", ``${\rm r}_B$", and $\gamma$  in the measurements
as when dealing with low statistical samples and low sensitivities
to direct {\it CP} violation effects due to small ``${\rm r}_B$"
values close to zero, non Gaussian effects and biaises arise when
fitting for the amplitude $\Gamma_\mp(m^2_-,m^2_+)$. These are
also natural parameters to describe the amplitude of the decay. A
simultaneous fit both to the $B^\pm$ decays and \Dztilde~$\to$
\KS\pim\pip\ decays is then performed to extract 12 parameters: $
\left\{ {\rm x}_\pm,   {\rm y}_\pm \right\}$ from \btodtildekpm, $
\left\{     {\rm x}^{*}_\pm,   {\rm y}^{*}_\pm \right\}$ from
\btodsttildekpm, and $\left\{  {\rm x}_{s\pm}, {\rm y}_{s\pm}
\right\}$ from \btodtildekstpm. In the last case,
\babar~\cite{BaBarGGSZ} deals with $($\KS\pimp$)_{{\rm non}-K^*}$
contributions, by defining an effective dilution parameter
$\kappa$ following the concept of ``generalized" {\it Cartesian
coordinates}~\cite{GronauCC}: ${\rm x}_{s\pm}^2+  {\rm y}_{s\pm}
^2=\kappa^2 {\rm r}_{sB}^2$, with $0\leq \kappa\leq 1$.
Belle~\cite{BelleGGSZ} fits directly for ${\rm r}_{sB}$ and
adresses these effects in the systematic uncertainty budgets.

Since the measurement of $\gamma$ arises from the interference
term in $\Gamma_\mp(m^2_-,m^2_+) $, the uncertainty in the
knowledge of the complex form of ${\cal A}_{D}$ can lead to a
systematic uncertainty. The additional phase of the previous
amplitude varies over the Dalitz plane. One can remark that it is
an additional technical difficulty with respect to the {\it GLW}
and {\it ADS} methods where there is only the unknown constant
strong phases $\delta_B$ and possibly an other strong phase
$\delta_D$ for \Dztilde\ pure two-body decays. The extraction of
$\gamma$ relies then on a Dalitz model for the complex amplitude
${\cal A}_{D}$ as a function of $(m^2_-,m^2_+)$. Both
\babar~\cite{BaBarGGSZ} and Belle~\cite{BelleGGSZ} use ``home
made" isobar models~\cite{CLEOKropp}  with coherent sums of
Breit-Wigner amplitudes, where the three-body \KS \pim \pip\ decay
is supposed to proceed via quasi two-body decay amplitudes only.
Theses models includes also an additional non-resonant term (NR):

\begin{equation}
{\cal A}_{D}(m^2_-,m^2_+) = \Sigma_r a_r e^{i \phi_r} {\cal
A}_r(m^2_-,m^2_+)+a_{NR} e^{i \phi_{NR}}.
\end{equation}

To build their isobar models, the \babar\ and Belle collaborations
use very high statistics  flavor-tagged \Dz\ sample ($D^{\ast +}
\rightarrow D^0 \pi^+_s$, where $\pi_s$ is a low momentum pion)
selected from data $e^+ e^- \to c \bar{c}$ events (respectively
about $390 \times 10^3$ and $260 \times 10^3$ events). These
samples have an excellent purity, larger than 97~\%.

The \babar\ model is based on 16 resonances (including three DCSD)
and on one NR term. Most of the resonance parameters are extracted
from the PDG~\cite{pdg2006}, except the $K^*_0(1430)$ that is
taken from the E791 experiment that uses an isobar model, while
the PDG quotes LASS parametrization. It includes also two ``ad
hoc" $\sigma(500)$ and $\sigma^{\prime}(1000)$ resonances to
describe the broad $\pi\pi$ $S$ waves. Their parameters are
determined in an effective way directly from the continuum \Dz\
data sample. In a second model (hereafter referred as the $\pi\pi$
$S$ wave K-matrix model) the treatment of the $\pi\pi$ $S$ wave
states in \Dz $\to$ \KS \pim \pip\ uses a K-matrix formalism to
account for the non-trivial dynamics due to the presence of broad
and overlapping resonances. This model is used in the Dalitz-model
systematic uncertainty determination. The total amplitude fit
fraction for the reference isobar model is about 1.2 and the value
of $\chi^2/n.d.o.f.$ of the fitted model is equal to 1.2.

The Belle model is built with 18 resonances (including 5 DCSD) and
one NR term, most of the resonance parameters are extracted from
the PDG~\cite{pdg2006}, except also the two ``ad hoc"
$\sigma(500)$ and $\sigma^{\prime}(1000)$ resonances, fitted
directly on the data. The total amplitude fit fraction is about
1.2 and the value of $\chi^2/n.d.o.f.$ of the fitted model is
equal to about 2.7.

A simultaneous fit to the $m_{ES}$ (or $m_{bc}$), $\Delta E$,
Fisher (for $B$ signal to $q\bar{q}$ light quarks separation)
variables and Dalitz model is then performed to extract the values
of the {\it Cartesian coordinates}, after selection of the $B^\pm$
candidates. The measurements have been performed by
\babar~\cite{BaBarGGSZ} and Belle~\cite{BelleGGSZ} with dataset
corresponding to $347 \times 10^6$ and $386 \times 10^6$ \BB\
pairs respectively. For the \btodtildekstpm\ mode \babar\ uses a
sample of $227 \times 10^6$ \BB\ pairs. Belle fits $331\pm17$
\btodtildekpm, $81\pm11$ \btodsttildekpm, and $54\pm8$
\btodtildekstpm, with purities respectively equal to 67~\%, 77~\%,
and 65~\%. The \babar\ $B^\pm$ sample corresponds to $398\pm23$
\btodtildekpm, $97\pm13$ ($93\pm12$) \btodsttildekpm, where
$D^{*0} \to D^0\pi^0$ ($D^0 \gamma$ (different from Belle)), and
$42\pm8$ \btodtildekstpm, and with similar purities as those from
Belle. The \babar\ dataset is therefore slightly larger than that
of Belle, due to better selection efficiencies.

\begin{figure}[h]
\centering
\includegraphics[width=80mm]{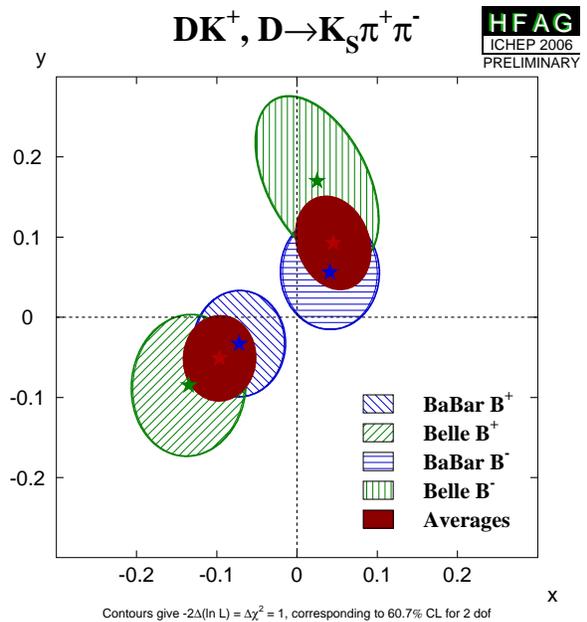}
\caption{World HFAG~\cite{HFAG} compilation on the
$ \left\{ {\rm x}_\pm,   {\rm y}_\pm
\right\}$  observables.} \label{fig:GGSZDK}
\end{figure}

\begin{figure}[h]
\centering
\includegraphics[width=80mm]{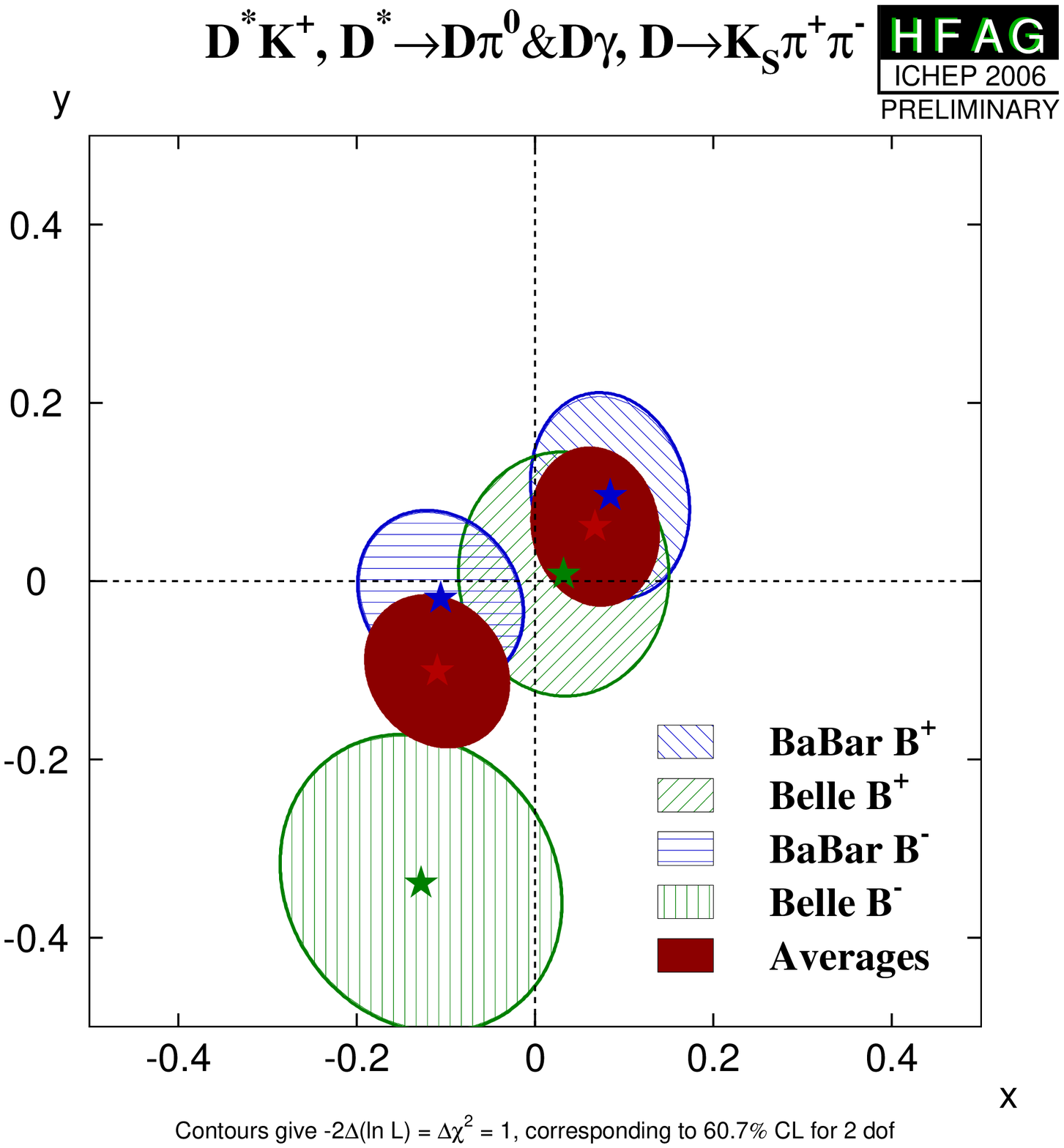}
\caption{World HFAG~\cite{HFAG} compilation on the
 $ \left\{ {\rm x}^*_\pm,   {\rm y}^*_\pm \right\}$ observables.}
\label{fig:GGSZDstarK}
\end{figure}

\begin{figure}[h]
\centering
\includegraphics[width=80mm]{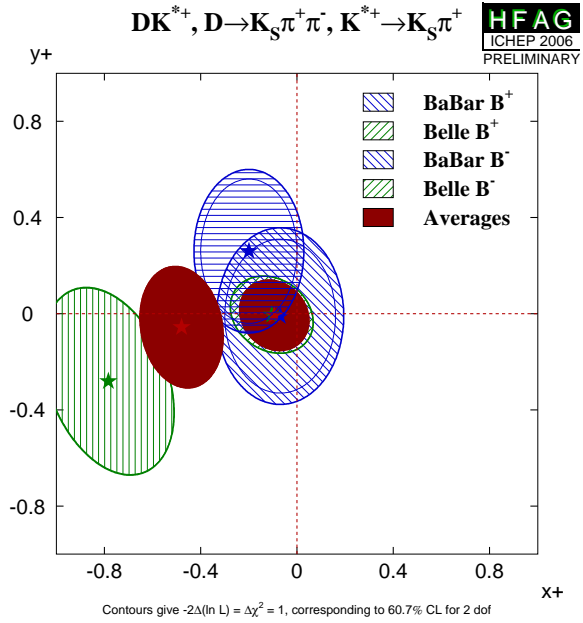}
\caption{World HFAG~\cite{HFAG} compilation on the
$ \left\{ {\rm x}_{s\pm},   {\rm y}_{s\pm} \right\}$  observables.} \label{fig:GGSZDKstar}
\end{figure}

On Figures~\ref{fig:GGSZDK}, \ref{fig:GGSZDstarK}, and
\ref{fig:GGSZDKstar} one can see the contours plots for the {\it
Cartesian coordinates} and for the three decay channels computed
by the HFAG collaboration~\cite{HFAG} from the measurements of the
two $B$ factories. These contours at 1 standard deviation do not
include the Dalitz  model uncertainty. An overall good agreement
between \babar\ and Belle measurements is visible. The precision
on the ${\rm x}_\pm$ variables is equivalent with what what
discussed in Sec.~\ref{section:GLW} with the {\it GLW} method. On
these Figures, one can see that the $B^+$ and $B^-$ contours are
flipped going from the \btodtildekpm\ decays to \btodsttildekpm,
because they differ by a relative phase $\pi$. The deviations from
the $(0,0)$ coordinates size the importance of the direct {\it CP}
violations effect and are proportional to the ``${\rm r}_B$"
nuisance parameters, possibly different in the measurement from
$B^+$ and $B^-$ candidates. Finally, the angle in between the
directions of the segments $[(0,0);({\rm x}_+,{\rm y}_+)]$ and
$[(0,0);({\rm x}_-,{\rm y}_-)]$ is by definition equal to
$2\gamma$.

At the end of the analysis, the 7 parameters: $\gamma$,
$\delta_B$, $\delta^*_B$, $\delta_{sB}$, ${\rm r}_B$, ${\rm
r}^*_B$, and $(\kappa .){\rm r}_{sB}$, are extracted from the 12
{\it Cartesian coordinates} using a frequentist approach that
defines a $5$-dimension $(D)$ (3-$D$) classical Neyman Confidence
Region (C.R.), in the case of \babar. The statistical extraction
method used by Belle is slightly different. A 7-$D$ C.R. is
computed using the refined frequenstist Feldman-Cousins ordering
technique in order to address the issue raised by possible
un-physical different values of ``${\rm r}_B$" as obtained for the
$B^+$ and $B^-$ populations.

The overall value for the {\it EW} {\it CP} phase measured by
\babar~\cite{BaBarGGSZ} is: $\gamma=[92 \pm 41 \pm 11 \pm
12]^\circ$ (the first uncertainty is statistical, the second
accounts for experimental systematic effects, and the third for
the {\it Dalitz model}), where it can be noticed that the
uncertainty coming from the employed Dalitz model would limit the
measurement at infinite statistics. This value is obtained with
the \btodtildekpm\ and \btodsttildekpm\ decays alone, and is
somewhat less precise than the 2005 published measurement. It is
due to lower estimated values for the ``${\rm r}_B$": ${\rm
r}_B<0.142$ $(0.198)$ and $0.016<{\rm r}^*_B<0.206$ $(0.282)$, at
1 (2) standard deviation(s), when the previous values were
respectively $0.12\pm 0.08 \pm 0.03 \pm 0.04 $ and $0.17 \pm 0.10
\pm 0.03 \pm 0.03$. As the sensitivity to $\gamma$ and as well the
precision of the measurement, varies as $1/{\rm r}_B$, so that
these smaller values explain a larger statistical error on
$\gamma$. No constraint at 1 standard deviation is derived from
the \btodtildekstpm\ mode alone, and a relatively loose upper
limit on $\kappa.{\rm r}_{sB}$ is set at 0.5.

The measurement performed by Belle has a better precision as the
estimated ``${\rm r}_B$" values are found to be larger: ${\rm
r}_B=0.159^{+0.054}_{-0.050} \pm 0.012 \pm 0.049$, ${\rm
r}^*_B=0.175^{+0.108}_{-0.099} \pm 0.013 \pm 0.049$, and ${\rm
r}_{sB}=0.564^{+0.216}_{-0.155} \pm 0.041 \pm 0.084$. They
correspond to a global value: $\gamma=[53^{+15}_{-18} \pm 3 \pm
9]^\circ$ ($8^\circ < \gamma < 111^\circ$ at 2 standard
deviations). This value is somewhat more precise from what one
would expect by scaling the statistics from the previously
published measurement in 2004. The value of ${\rm r}^*_B$ has been
shifted up, as it was previously equal to $0.12^{+0.16}_{-0.11}
\pm 0.02 \pm 0.04$ and ${\rm r}_B$ did not change significantly as
it was previously $0.21 \pm 0.08 \pm 0.03 \pm 0.04$. The
experimental systematic uncertainty has been significantly reduced
due to the use of a control high statistic control samples of
$D^{(*)0} \pi^-$ and $D^{*-}\pi^+$ $B$ decays.

More recently \babar~\cite{BaBar3piDalitz} has studied the Dalitz
decay \btodtildekpm, where the \Dztilde\ decays to \piz\pim\pip.
With respect to the \KS \pim \pip\ mode the signal rate is divided
by about a factor 2. The background is also larger due to the
presence of a \piz\ meson. \babar\ has measured for the first time
the isobar model of this Dalitz decay. Due to significant
non-linear correlations it has been found that with the present
statistic the {\it Cartesian coordinates} can not be used in the
global fit extraction, neither the $\gamma$, ${\rm r}_B$, and
$\delta_B$ constants. If one defines ${\rm z}_\pm \equiv {\rm r}_B
e^{i(\delta_B \pm \gamma)}$, it has been demonstrated that the fit
biases are strongly reduced when fitting for the so-called {\it
polar coordinates}: $\rho_\pm\equiv \vert {\rm z}_\pm - {\rm x}_0
\vert$ and $\theta_\pm \equiv \tan^{-1}{ \left( \frac{{\rm
y}_\pm}{{\rm x}_\pm -{\rm x}_0} \right)}$, where $x_0$ is a
coordinate transformation parameter equal to 0.85. For no {\it CP}
violation the $B^+$ and $B^-$ contours are centered in the
$(\rho_\pm, \theta_\pm)$ plane on the coordinate $(0.85,
180^\circ)$. \babar\ obtains: $\rho_-=0.72 \pm 0.11 \pm 0.04$,
$\rho_+=0.75 \pm 0.11 \pm 0.04$, $\theta_-=(173 \pm 42\pm
2)^\circ$, and $\theta_+=(147 \pm 23 \pm 1)^\circ$ (where the
uncertainties are respectively for statistical and for the
systematic effects). So far no strong deviation from that position
as been established and no attempt has been made to extract the
physics constants $\gamma$, ${\rm r}_B$, and $\delta_B$ by \babar.
A quite loose constraint on $\gamma$ is foreseen.

\section{Measuring $\gamma$ with neutral $B$ decays}

The decays of neutral $B$ mesons that allow to constrain
$\sin(2\beta+\gamma)$ have either small sensitivity to the
$V_{ub}$ phase or small branching fractions. At the present time
these measurements can only help in improving the overall picture
on the determination of the CKM-angle $\gamma$. The presence of
hadronic parameters in the observables (${\rm r}$ and $\delta$,
the amplitude ratio and the strong phase difference between the
$V_{cb}$ and $V_{ub}$  interfering amplitudes), as for previously
discussed for charged $B$ decay, complicates the extraction of the
weak phase informations. The problem here is even more crucial as
in case of $D^{(*)\pm}\pi^{\mp}$ and $D^{\pm}\rho^{\mp}$ mode the
$V_{ub}$ amplitude counts only as about 2~\%~\cite{sin2BplusG} of
the total amplitude and in the case of \btodtildekstzero\ these
modes are so rare that the value of ${\rm r}$ can not yet be
directly measured. There exist however approaches based on SU(3)
symmetry to estimate the magnitude of the nuisance parameter ${\rm
r}$ and as a consequence to set constraints on
$\sin(2\beta+\gamma)$.

\subsection{{\it CP} asymmetry in neutral $D^{(*)\pm}\pi^{\mp}$ and $D^{\pm}\rho^{\mp}$ $B$ decays}

This technique has been proposed~\cite{sin2BplusG} as the
$D^{(*)-}\pi^{+}$ and $D^{-}\rho^{+}$ decays can be produced
either in the decay of a \Bz\  or a \Bzb, respectively through the
$\bar{b} \to \bar{c} u\bar{d}$ (the ``$V_{cb}$ amplitude", Cabibbo
favored (${\cal A }_{CF}$)) or $b \to u \bar{c}d$ transition (the
``$V_{ub}$ amplitude", doubly Cabibbo suppressed (${\cal A
}_{DCS}$)). These are pure tree decays with relatively large
branching ratios of the order of $0.3-0.8\%$. Their relative weak
phase is $\gamma$. An additional weak phase $2\beta$ may come from
initial \Bz-\Bzb\ mixing. Another unknown relative strong phase
$\delta$ arises from strong interaction in the final state in
between these two amplitudes. Due to mixing, the
$D^{(*)\pm}\pi^{\mp}$ decay rate evolves with time as:

\begin{equation}
\begin{array}{rcl}
P(B^0 \to D^{(*)\pm}\pi^{\mp},\Delta t) & \propto & 1 \pm
C^{(*)}\cos(\Delta m_d \Delta t) \nonumber \\
& + & S^{(*) \mp}  \sin(\Delta m_d \Delta t), \\

P(\kern 0.18em\overline{\kern -0.18em B}^0 \to
D^{(*)\mp}\pi^{\pm},\Delta t) & \propto &   1 \mp
C^{(*)}\cos(\Delta m_d \Delta t) \nonumber \\
& - & S^{(*) \pm} \sin(\Delta m_d \Delta t),
\end{array}
\end{equation}

where $\Delta m_d$ is the mixing frequency, and $\Delta t$ is the
time difference in between the time of the $B \to
D^{(*)\pm}\pi^{\mp}$ decay (hereafter referred as the
reconstructed $B$ meson, $B_{rec}$) and the decay of the other $B$
meson (hereafter referred as the tagging $B$ meson, $B_{tag}$). In
the above equation the flavor \Bz\ (\Bzb) can be experimentally
determined from the flavor of the $B_{tag}$ (using a flavor
specific final state). Therefore the technique employed to extract
the relevant constants is similar to the time dependent analyses
performed with $(c\bar{c})$\KS\ decays for the determination of
$\sin(2\beta)$~\cite{ChHCheng}. The parameter $C^{(*)}$ and
$S^{(*) \pm}$ are given by:

\begin{equation}
C^{(*)} \equiv \frac{1-{\rm r}^{(*)2}}{1+{\rm r}^{(*)2}}(\simeq
1),
\end{equation}

\begin{equation}
S^{(*) \pm} \equiv \frac{2{\rm r}^{(*)}}{1+{\rm r}^{(*)2}}\sin(2
\beta + \gamma \pm \delta^{(*)}).
\end{equation}

We can also define the strong phase differences in between ${\cal
A }_{CF}$ and  ${\cal A }_{DCS}$ as $\delta^{(*)}$ and ${\rm
r}^{(*)}$ as the ratio $\left\vert {{\cal A }_{CF}}/{{\cal A
}_{DCS}} \right \vert$. There exist as well two constants ``${\rm
r}$" and ``$\delta$" for the $D^{\pm}\rho^{\mp}$ mode. As expected
from the DCSD phenomenon~\cite{sin2BplusG}, the ``${\rm r}$"
constants are expected to be of the order of~2~\%. Due to the
small value of ${{\rm r}^{(*)2}}(\sim 10^{-4}$) it obviously
impossible to extract ``${\rm r}$"  from the measurements  with
the present statistic of the available samples (i.e., from
$C^{(*)}$). It is also mandatory to have large data sample for
extracting statistically significant measurements of $S^{(*)
\pm}$.

The \babar~\cite{BaBarsin2BplusG} and Belle~\cite{Bellesin2BplusG}
collaborations are using two different techniques to select
high-statistics samples of $D^{(*)\pm}\pi^{\mp}$ and
$D^{\pm}\rho^{\mp}$ $B$ decays. These measurements have been
obtained with dataset corresponding respectively to $232 \times
10^6$ and $386 \times 10^6$ \BB\ pairs.  A full reconstruction
technique  is accessible for the three above decays, while a
partial reconstruction technique allows to use the $D^*\pi$  decay
mode only.

For the exclusive method, the full decay chain is reconstructed.
Therefore an excellent purity of the order of 90~\% for $D^*\pi$
decay is achieved. The price to pay, in addition to the limited
tagging efficiency and when determining the $B_{tag}$ flavor, is
the relatively limited event yields. Belle gets about $30 \times
10^3$ events for the $D^*\pi$ decay.

The other method is based on a partial reconstruction where only
the soft (low momentum) $\pi$ track of the $D^*$-meson decay  and
the prompt (high momentum) $\pi$ track of the $B$ decay are
detected. As the $D$'s are not explicitly reconstructed one gains
on the secondary decay branching ratios. The price to pay is a
relatively lower purity that depends on the tagging category. It
goes from 30~\% in case of kaons tags to about 55~\% for prompt
leptons tags. The statistics for this method can be enhanced up to
5 or 6 times with respect to the one obtained for the fully
exclusive technique. With $232 \times 10^6$ \BB\ pairs, \babar\
analyse a data sample of about $71 \times 10^3$ $K$-tagged
$D^*\pi$ ($19 \times 10^3$ lepton-tagged). The $K$-tagged sample
has about 4 times the statistics of the leptons-tagged one.

An important experimental difficulty has to be mentioned. As the
expected {\it CP} asymmetries for these measurements are small,
the interferences of $b \to u$ and $b \to c$ amplitudes in the
decay of the $B_{tag}$ have to be taken into account. They dilute
the effective $B_{rec}$ {\it CP} asymmetry. The \babar\
collaboration uses an alternative parametrization to the $S^{(*)
\pm}$:

\begin{equation}
\begin{array}{rcl}
 a & \equiv &  2{\rm r} \sin(2\beta+\gamma)\cos(\delta),\\
 b & \equiv &  2{\rm r}^{\prime} \sin(2\beta+\gamma)\cos(\delta^{\prime}),\\
 c & \equiv &  2 \cos(2\beta+\gamma)({\rm r} \sin(\delta)-{\rm
 r}^{\prime} \sin(\delta^{\prime})).
\end{array}
\end{equation}

For each $D^{(*)\pm}\pi^{\mp}$ and $D^{\pm}\rho^{\mp}$ decay mode,
each measurement technique (partial or full reconstruction) and
for each tagging category one gets a different interference on the
tagging side, as the background differ from one case to the other.
Effective ``${\rm r}^\prime$" and ``$\delta^\prime$" parameters
are therefore derived. One notes that ${\rm r}^\prime$
automatically vanishes in $a$ and $c$ when using the lepton
tagging category.

\babar~\cite{BaBarsin2BplusG} uses both kind of tags for the
partial reconstruction technique and and only lepton tags for the
full reconstruction technique. Belle~\cite{Bellesin2BplusG} uses
only lepton tags for the partial reconstruction method (thus
limiting the yield of useful $D^*\pi$ candidates for the {\it CP}
violation measurement), and all tags categories for the fully
exclusive method. The DCSD interferences in the $B_{tag}$ side are
measured using control samples of $D^{(*)} l \nu$ $B$ decays.

\begin{figure}[h]
\centering
\includegraphics[width=80mm]{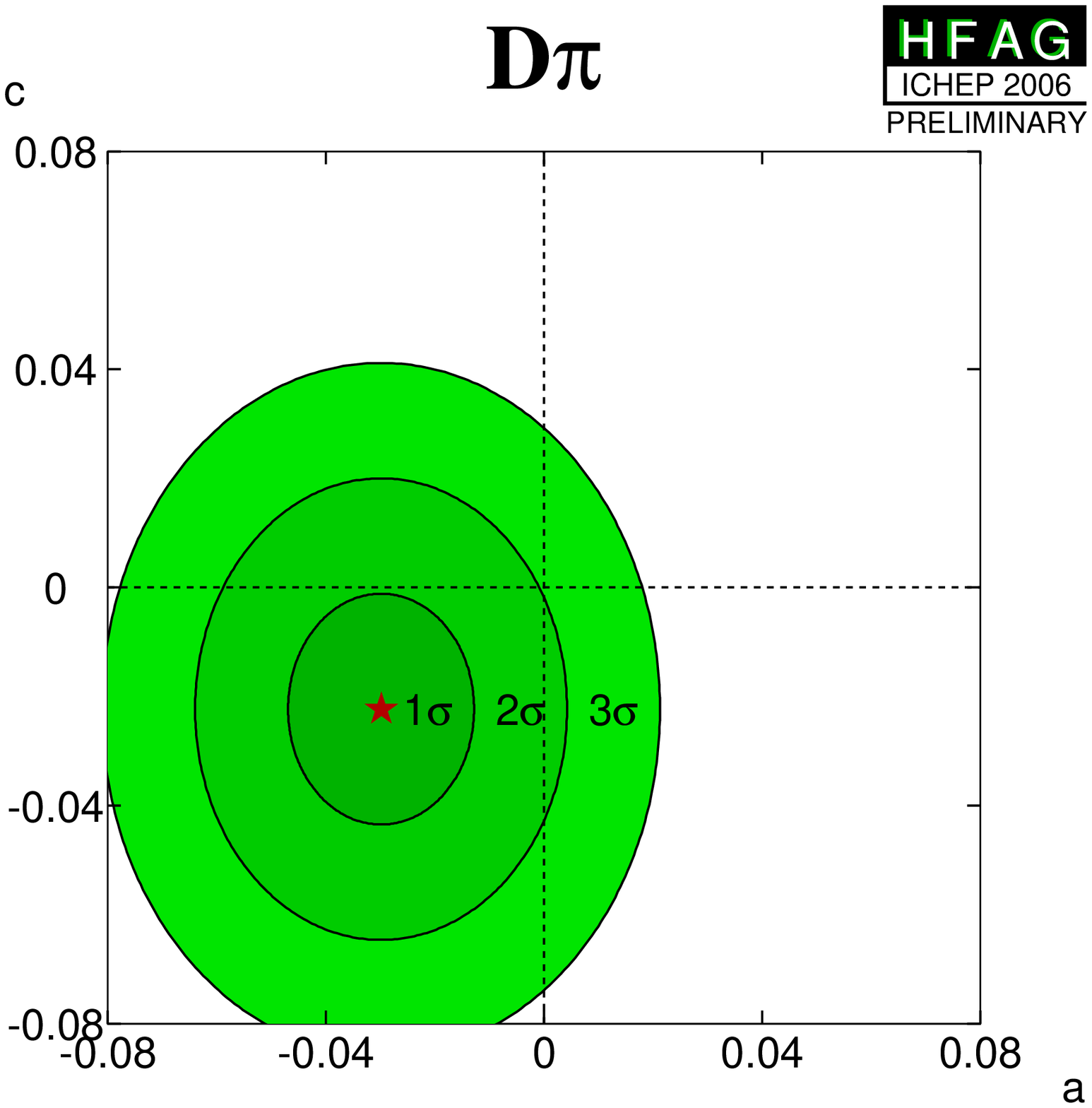}
\caption{Graphical view of the world HFAG~\cite{HFAG}
compilation for the $a$ and $c$ parameters for the
$D^{\pm}\pi^{\mp}$ decay mode.} \label{fig:acD}
\end{figure}

\begin{figure}[h]
\centering
\includegraphics[width=80mm]{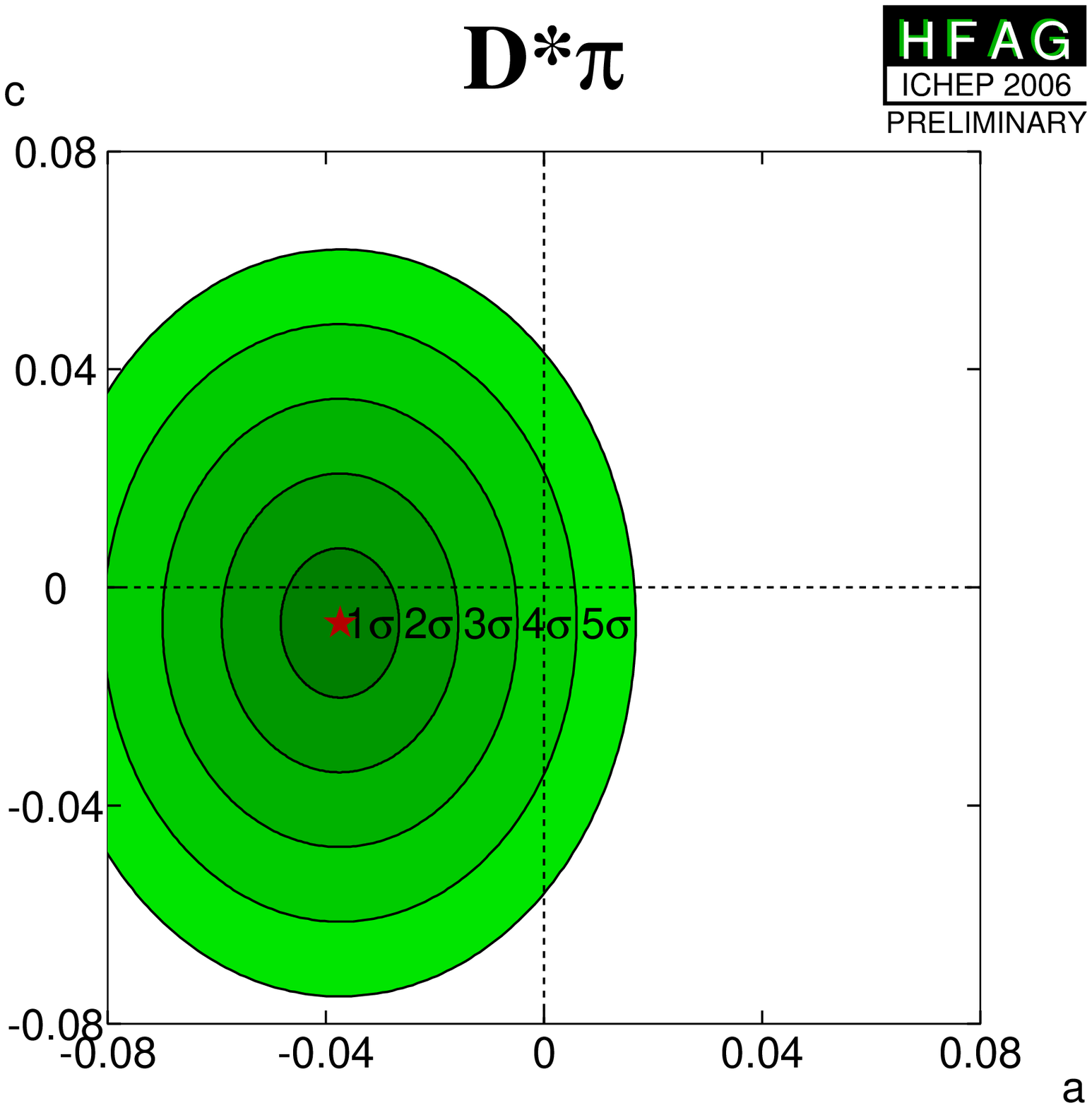}
\caption{Graphical view of the world HFAG~\cite{HFAG}
compilation for the $a^*$ and $c^*$ parameters for the
$D^{*\pm}\pi^{\mp}$ decay mode.} \label{fig:acDstar}
\end{figure}

These measured values for $a$ and $c$, in the two methods and for
the three types of decay mode, are in good agreement in between
the two experiments, the detailed values can be found in the HFAG
collaboration winter 2007 document~\cite{HFAG}. The combined
values of $(a;c)$ are: $(-0.030 \pm 0.017;-0.022 \pm 0.021)$ for
$D^{\pm}\pi^{\mp}$, $(-0.037 \pm 0.011;-0.006 \pm 0.014)$ for
$D^{*\pm}\pi^{\mp}$, and $(-0.024 \pm 0.033;-0.098 \pm 0.058)$ for
$D^{\pm}\rho^{\mp}$. Figures~\ref{fig:acD} and \ref{fig:acDstar}
show in a graphical way the combined $(a;c)$ values for
$D^{\pm}\pi^{\mp}$ and $D^{*\pm}\pi^{\mp}$.

A 3.4 $\sigma$ deviation from 0 is visible for $a$ in the $D^*$
mode indicating that observation of {\it CP} violation is within
reach for the $B$ factories with some additional statistics. The
currently published results by \babar\ and Belle are anyway not
using the full available dataset.

Both \babar~\cite{BaBarsin2BplusG} and
Belle~\cite{Bellesin2BplusG} have extracted limits on $\vert
\sin(2\beta+\gamma) \vert$. For this it is mandatory to estimate
the value of ``${\rm r}$". This is done using SU(3) flavor
symmetry, with available branching fractions (including recent
\babar~\cite{BaBarDspi} measurement for the $D^{(*)+}_s \pi^-$
modes), and lattice calculations for decay constants. These
extractions of ${\rm r}$ have a relative 30~\% uncertainty due to
estimation of SU(3) breaking and limited knowledge on $W$-exchange
and annihilation diagrams. With all the three \babar\ sets lower
limits on $\vert \sin(2\beta+\gamma) \vert$ respectively equal to
$0.64 \ (0.40)$ at 68~\% of C.L. (90~\%). With the
$D^{*\pm}\pi^{\mp}$ mode  Belle sets the lower limit $0.44 \
(0.13)$ at 68~\% of C.L. (95~\%). With the $D^{\pm}\pi^{\mp}$
Belle sets lower limit  $0.52 \ (0.07)$ at 68~\% of C.L. (95~\%).

\begin{figure}[h]
\centering
\includegraphics[width=80mm]{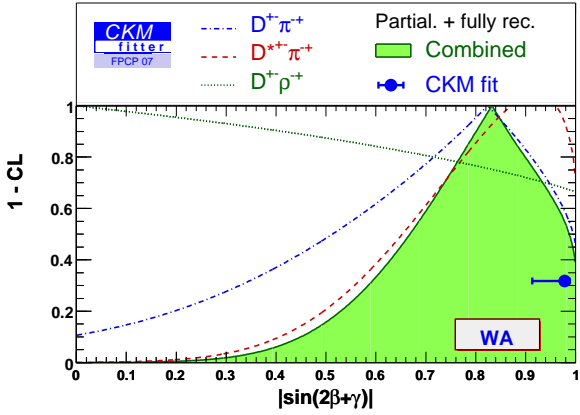}
\caption{1-C.L. constraint profile as obtained for $\vert
\sin(2\beta+\gamma) \vert$~\cite{CKMFitter}.}
\label{fig:Sin2BplusGCKM}
\end{figure}

Based on an updated and more sophisticated version of the \babar\
model, a re-scattering SU(3) symmetry model~\cite{Baak}, the
following values of ``${\rm r}$" are computed: ${\rm
r}^{D\pi}=(1.53 \pm 0.33 \pm 0.08)\%$, ${\rm r}^{D^*\pi}=(2.10 \pm
0.47 \pm 0.11)\%$, and ${\rm r}^{D\rho}=(0.31 \pm 0.59 \pm
0.02)\%$, where the first uncertainty is a Gaussian error for
SU(3) breaking from non-factorizable contributions and the second
is a 5~\% flat error for SU(3) breaking from $W$-exchange and
annihilation diagrams. A global combined
frequentist~\cite{CKMFitter} constraint of all the available
results is shown for $\vert \sin(2\beta+\gamma) \vert$  on
Figure~\ref{fig:Sin2BplusGCKM}, where  the C.L. distributions are
displayed. A lower limit on this quantity is set at $0.59 \
(0.37)$ at 68~\% of C.L. (95.5~\%).

\subsection{{\it CP} asymmetry in neutral \btodtildekstzero\ $B$ decays}

It has been proposed~\cite{GLW,ADS,DZKz} that the rare neutral
\btodtildekstzero\ $B$ decays can be used for the time dependent
{\it CP} asymmetry measurement of $\sin(2 \beta+ \gamma)$. These
final states can be produced through $b \to c$ or $\bar{b} \to
\bar{u}$ transitions. Both are color-suppressed and
Cabbibo-suppressed transitions and are of the same order of
magnitude ($\propto \lambda^3$). Despite the rareness of theses
modes, it has been stressed that the ratio of the ratio
$\tilde{\rm r}_B$ should be relatively large and of the order of
0.4, thus making these modes appealing with large $B$ meson
dataset. As $\tilde{\rm r}_B$ is large enough, in the time
dependent analysis and as opposed to the previously described
neutral $B$ decays, it is feasible to extract both $S^{(*)\pm}$
and $C^{(*)}$ coefficients, so that no theoretical assumption nor
any model is needed to measure $\sin(2 \beta+ \gamma)$, ${\tilde
\delta}_B$, and $\tilde{\rm r}_B$ at the same time.

So far, Belle with $88 \times 10^6$ \BB\ pairs and
\babar~\cite{DzKZBaBarBelle} with $226\times 10^6$ \BB\ pairs,
have measured the branching ratios of these decays. They lie in
the range $4-5\times 10^{-5}$, while the mode $\bar{B}^0 \to
\bar{D}^0 \bar{K}^{*0}$ has not yet been observed. The most
precise upper limit for its branching fraction is $1.1\times
10^{-5}$ at 90~\% C.L.. It is then  obvious that no direct
measurement of $\tilde{\rm r}_B$ is yet accessible.

Using the self tagging decay $\bar{K}^{*0} \to K^-\pi^+$, it is
possible to estimate the ratio $\vert {\cal A }(\kern
0.18em\overline{\kern -0.18em B}^0 \to \bar{D}^0 \bar{K}^{*0}) /
{\cal A }(\kern 0.18em\overline{\kern -0.18em B}^0 \to {D}^0
\bar{K}^{*0}) \vert$. These decays can be therefore distinguished
by the kaons charges correlations in the $D$ and $K^*$ meson
decays (the $V_{ub}$ transition would have opposite charges kaons,
as for the {\it ADS} method described in Sec.~\ref{section:ADS}).
Again the sensitivity to $\tilde{\rm r}_B$ is diluted by the
presence of DCS decays: $D^0 \to K^+\pi^-$, $K^+\pi^-\pi^0$, and
$K^+$ \pim \pip \pim\ modes for which ${\rm r}_D$ constants have
measured relatively precisely~\cite{pdg2006}. From the ratio of
branching ratios ${\bf R}=\Gamma(\kern 0.18em\overline{\kern
-0.18em B}^0 \to (K^+X_i^-)_D \bar{K}^{*0})/\Gamma(\kern
0.18em\overline{\kern -0.18em B}^0 \to (K^-X_i^+)_D
\bar{K}^{*0})$, where $X^\pm_i=$\pipm, \pipm\piz, or \pipm \pim
\pip, one can extract constraints on $\tilde{\rm  r}^2_B$, when
knowing ${ r^2_D}$ and doing assumptions on ${ \tilde{\delta}_B}$,
$\gamma$, and ${ \delta_D}$ (see Sec.~\ref{section:ADS}). Doing
this, \babar~\cite{DzKZBaBarBelle} has set the upper limit
$\tilde{ \rm r}^2_B<0.40$ at 90~\% of C.L.. This suggests that, as
the branching fraction of the decay $\bar{B}^0 \to \bar{D}^0
\bar{K}^{*0}$ is still unknown, that this technique is still not
yet powerful enough with the existing $B$ factories dataset and
will not even be in near future.

\section{Conclusions, perspectives, and global constraints on
$\gamma$}

We have presented a review on the measurements of the CKM-angle
$\gamma$ ($\phi_3$) as performed at the $B$ factories PEP-II and
KEK$B$ by \babar\ and Belle collaborations. These measurements
were considered as more or less impossible for these \FourS\
experiments a few years ago. Thought we have not yet entered the
era of precision, many methods using charged or neutral $B$-decays
have been employed, pioneering the measurements at the future LHCb
experiment, or possibly at future super $B$ factories.

Before to get there, the present existing $B$ factories will have
to update these results with already existing additional dataset
and non negligible forthcoming statistics. In addition, many
refinements and new methods continue to be
developed~\cite{CKM2006}. One can therefore anticipate substantial
improvements. All the machinery is in place, but all these
measurements are by far dominated by statistics uncertainty.
Puzzles remain such as the exact value of the ${\rm r}_B$ that
sizes the sensitivity to $\gamma$ and that is  always exploited in
each of the previously described methods. The bigger it is, the
shorter will be the way to precision era.

So far the {\it GGSZ} method exploiting the Dalitz Decay
\Dztilde$\to$\KS \pim \pip\ in charged $B$ decays
$\tilde{D}^{(*)0}K^{(*)-}$ continues to provide us with the most
powerful constraint on $\gamma$. The Dalitz model systematic
uncertainty is not yet a concern at the present time.
Nevertheless, it is possible to reduce the model dependence
systematic uncertainty by using the original idea of the GGSZ
method~\cite{ADS,GGSZ}. Using CLEO-c and future $\tau$-charm
factories it is possible~\cite{AsnerBondarPoluektov} to produce at
the $\psi(3770)$ resonance coherent states of \Dz\ and \Dzb\
pairs. The amplitude $\Gamma_\mp(m^2_-,m^2_+)$ is no more fitted
but replaced by event yields of {\it CP} and flavor tagged
$\tilde{D}^0$ within bins of the Dalitz plan. With the existing
CLEO-c statistics (280~\invpb\ corresponding to about 570 {\it CP}
tagged events) this method should help reducing this uncertainty
down to $6^\circ -7^\circ$. For a projection to about 750~\invpb,
corresponding to about 1500 {\it CP} tagged events, one should
reduce it down to about $4^\circ$. At a Super $B$ factory with
50~\invab, in addition to 10~\invfb\ of data collected at future
$\tau$-charm factories, on should ultimately be able to measure
$\gamma$, in a model-independent way, with an accuracy of the
order of $2^\circ$.

\begin{figure}[h]
\centering
\includegraphics[width=80mm]{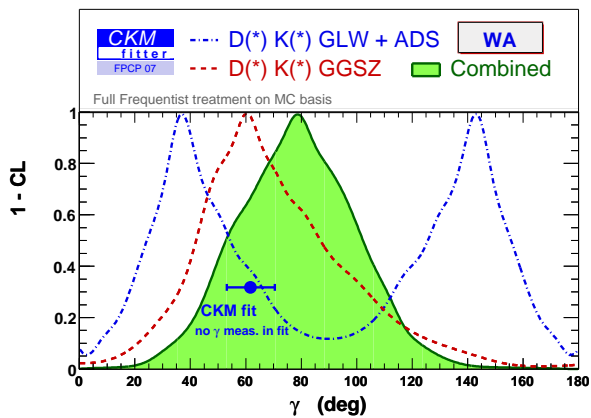}
\caption{1-C.L. constraint profile as obtained for
$\gamma$~\cite{CKMFitter}.} \label{fig:GammaCKM}
\end{figure}

To conclude, we present on Figure~\ref{fig:GammaCKM} the
combination on the constraints on $\gamma$ obtained with the
charged $B$ decays. The present fit~\cite{CKMFitter} is performed
with the HFAG~\cite{HFAG} combinations of 32 observables (the
\Rcppm, \Acppm, \RADS, and {\it Cartesian coordinates}) to extract
11 physics constants (we have evaluated the strong phases and
relative amplitudes ratios for $\tilde{D}^0$ decays to $K\pi$ and
$K\pi\pi^0$ modes). The value $\gamma=(77\pm 31)^{\circ}$ is
extracted. This is coherent with the 1~$\sigma$ interval from a
global CKM coherence fit where these measurement are absent:
$52.8^\circ < \gamma < 70.1^\circ$. At 90~\% of C.L. one gets from
that fit: ${\rm r}^{DK}_B<0.13$, ${\rm r}^{D^*K}_B<0.13$, and
${\rm r}^{DK^*}_B<0.27$. Adding the information coming from the
constraint on $\vert \sin(2\beta+\gamma) \vert$ a slightly more
precise value is obtained: $\gamma=(78^{+19}_{-26})^{\circ}$.

\begin{acknowledgments}
I would like to thank the organizers and especially Peter Krizan
and his team, for the choice of the beautiful location of Bled
lake for that conference. The richness of the scientific program
and the perfection of the organization and the anticipation of
every single practical detail was quite impressive.

I would like also to thank the \babar\ and Belle physics groups
conveners:  Jean-Pierre Lees, Matteo Rama, and Karim Trabelsi for
the many fruitful discussions we had when I prepared that talk.
Finally, I would like to thank my CKMfitter colleagues, and
especially Heiko Lacker, again Karim Trabelsi, and St\'ephane
T'Jampens for their help when updating the global constraints on
$\gamma$ for that conference.

\end{acknowledgments}

\bigskip 

\end{document}